# How Peripheral Interactive Systems Can Support Teachers with Differentiated Instruction: Using FireFlies as a Probe

**Nine Sellier**
Department of Industrial Design
Eindhoven University of Technology
The Netherlands
n.sellier@student.tue.nl

**Pengcheng An**
Department of Industrial Design
Eindhoven University of Technology
The Netherlands
p.an@tue.nl

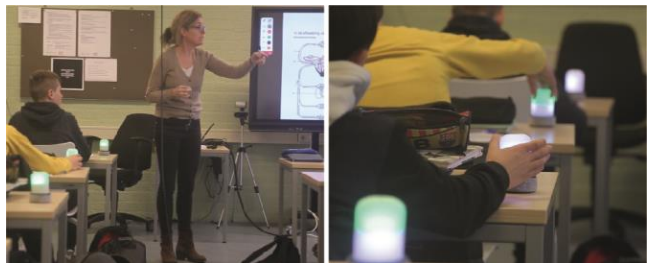

**Figure 1. The FireFlies probe situated in the classroom.**

### ABSTRACT
Teachers' response to the real-time needs of diverse learners in the classroom is important for each learner's success. Teachers who give differentiated instruction (DI) provide pertinent support to each student and acknowledge their differences in learning style and pace. However, due to the already complex and intensive routines in classrooms, it is demanding and time-consuming for teachers to implement DI on-the-spot. This study aims to explore how to ease teachers' classroom differentiation by enabling effortless, low-threshold student-teacher communications through a peripheral interactive system. Namely, we present a six-week study, in which we iteratively co-designed and field-tested interaction solutions with eight school teachers, using a set of distributed, interactive LED-objects (the 'FireFlies' platform). By connecting our findings to the theories of DI, we contribute empirical knowledge about the advantages and limitations of a peripheral interactive system in supporting DI. Taken together, we summarize concrete opportunities and recommendations for future design.

### CCS Concepts
•**Applied computing ~ Education ~ Interactive learning environments** • Human-centered computing ~ Human computer interaction (HCI) ~ Empirical studies in HCI

### INTRODUCTION
One of the most difficult challenges teachers face today is being able to respond to the diverse and varying needs of individual learners [54] in the classroom. Each learner differs in characteristics such as personalities, learning styles, and existing academic skills, and thus needs differentiated supports during class sessions [44]. Teachers' response to the real-time needs of each unique learner in the classroom is crucial for the leaner to grow. Teachers who give differentiated instruction (DI) [54, 59] provide specific alternatives for their students to learn as deeply and quickly as possible without assuming that their individual road maps for learning are identical to each other [54]. However, although being widely considered as an important pedagogical competence [9], differentiated instruction often remains a challenging task for teachers to perform on the spot, during classroom teaching. This is due to teachers' already complex and intensive routines in the classroom [17], which makes it cognitively demanding and time-consuming for teachers to implement DI on the fly.

Teachers' DI during classroom teaching can be supported by classroom orchestration [21] or teaching augmentation systems [7], which have been designed in various forms, such as real-time learning analytics dashboards [63, 52, 38], teacher wearables [32, 66], or distributed peripheral interactive systems [12, 64,m1]. For instance, real-time learning analytics dashboards display students' learning processes with learning software (e.g., [38]) to ease teachers' differentiation in blended classrooms. In both blended and face-to-face classrooms, distributed peripheral interactive systems [2, 6, 8, 64] can support teachers' differentiation by enabling minimalist, non-verbal student-teacher communication (e.g. of students' help-seeking or learning status), via distributed, interactive ambient lamps.

In this paper, we focus on exploring how distributed peripheral interactive systems can support teachers' DI in classroom teaching. Prior research in HCI [6, 8, 11, 12, 30, 64] has implied the effects of distributed peripheral interactive systems on particular aspects of DI (e.g., reducing students' waiting durations for receiving help from the teacher [1]). However, these empirical findings have rarely been connected to the established theories of DI from education science [54, 59, 61, 23, 56], in order to holistically inform future designers of the benefits and limitations of distributed peripheral interactive systems in supporting DI.

To tackle this unaddressed opportunity, in this study, we connect our design research practice to the theories of DI, to widely probe both the benefits and limitations of a peripheral interactive system in supporting teachers' DI during classroom teaching. To do so, we used an open-ended peripheral interactive system (FireFlies) as a *technology probe* [33], to co-design and evaluate different interaction solutions with eight teachers from a secondary school in the Netherlands, over six weeks. As such, this field study has combined both co-design need-finding and field evaluation, and resulted design insights that have been explicitly connected to the DI theories.

This paper thereby contributes (i) contextualized knowledge about the benefits and limitations of distributed peripheral interactive systems in supporting the different elements of teachers' DI in the classroom, and (ii) interaction design implications and recommendations to inform future practice.

**THEORIES OF DIFFERENTIATED INSTRUCTION**
Differentiation is the teachers' proactive response to the diverse and varying needs of their students [59]. In general, teachers can give DI by reaching out to an individual or a small student-group and by giving pertinent supports for each learner to learn as deeply and quickly as possible [59].

A body of literature from educational science has been focused on developing the theoretical accounts of differentiated instruction [54, 59, 61, 23, 56]. According to Tomlinson's [59] renowned conceptual model of Differentiation of Instruction, teachers can differentiate on the content, process, product and learning environment (i.e. the four classroom elements of DI) according to the , interests and learning profile of their students. However, findings in recent studies have shown the lack of credible evidence for the learning styles theory [35, 43] and showed that matching instruction to the strengths and preferences of the students does not necessarily guarantee academic success [43, 46, 53]. Because of these findings, Pham [44] emphasizes that teachers should refer to the students' readiness, which is their current proximity to a learning goal [61], rather than the student preferences or learning styles while differentiating. This implies that effective DI heavily relies on teachers' on-the-spot sensemaking of students' real-time needs during learning activities.

Despite the above arguable differences, there is a clear consensus among these DI theories upon the four major classroom elements a teacher can differentiate through: content, process, product, and learning environment (see Figure 1). In this study, we looked at how a distributed peripheral interactive system can support teachers in differentiating these four classroom elements.

*Content* is *"what the student needs to learn or how the student will get access to the information"* [57]. Teachers can differentiate the content during the lesson by continually evaluate students' understandings. This can result in teachers' responsive modifications of both interpersonal and whole-class instructions [54]. Teachers can, for example, change the level or pace of the instruction or offer multiple ways in which they present the information. Also, teachers can meet with small groups and give extended instruction to advanced learners or re-teach an idea or skill to struggling learners [57].

*Process* describes the *"activities in which the student engages in order to make sense of or master the content"*[57]. Important aspects of differentiating the process are to facilitate multiples learning activities for students to reach defined learning goals [20] and to allow students to learn at their own pace [57]. Teachers can differentiate the process by finding out where their students are in their learning processes and adaptively provide supports [34].

*Product* is the instrument *"through which students demonstrate and extend what they have learned"* [59]. Teachers can differentiate the product by diversifying the ways that the students are assessed, or by encouraging students to create their own product assignments [57].

*Learning environment* is the *"climate"* of the classroom [59]. A differentiated classroom respects learners' diversity, makes them feel safe to express their needs, and encourage their accountability and autonomy [58]. Learners should feel that they are listened to and be aware that their peers each learn differently [26]. Also, the classroom should support both individual and collaborative working [57]. And learners should expect that they will get help when teachers cannot help immediately [57].

**RELATED DESIGN CASES**
Prior HCI work has presented design cases about how technologies could support teachers in particular aspects related to DI. We address related cases in this section with a main focus on distributed peripheral interactive systems.

Various forms of technologies that aimed for classroom orchestration [21], or teaching augmentation [7], can benefit teachers' classroom differentiation. For instance, in blended (computer supported) classrooms, real-time learning analytics dashboards (e.g., MTDashboard [38], or SAM [63]) can help teachers to monitor students' learning processes taking place in learning software, and thereby modify interventions based on individual needs. Many of these dashboard interfaces also support teacher-student online communication, which further extends teachers' ability to orchestrate. On top of that, several emerging design cases explored wearable devices such as smartglasses [32, 66], which suggest benefits in seamlessly integrating real-time information to teachers' vision. Lumilo [30], for example, augment teachers' decision-making by showing information from AI-based learning analytics algorithms.

Motivated by the vision of clam technology [65], there are also design cases supporting teachers through peripheral [13] or ambient information [45] on wall-mounted, centralized displays (e.g. [3, 22]). Lernanto [3], for instance, translate learning analytics data into colour patterns on an LED panel, to subtly complement teachers' classroom differentiation. Similarly, Tinker Board [22] offers ambient information to support teachers' orchestration of CSCL sessions. While peripheral information can be shown through above centralized displays, it can also be conveyed through peripheral interactive devices distributed on student desks, such as tangibles [25] or ambient lamps [2, 6, 8, 64].

Our work has been focused on such distributed peripheral interactive systems. Lantern [2, 1], as a renowned example, consists of multiple ambient lamps that enable student-teams to communicate their status and help requests to university lecturers. Lantern proves to increase the efficiency of help-seeking processes. The FireFlies platform [64] resulted similar design cases in K-12 classrooms. FireFlies is a set of open-ended, programmable lamps, which afford both students' and teachers' manipulations, and can display various types of data through its upper and lower LED segments [8, 6]. A series of applications of FireFlies explored how to support particular aspects of teaching that can be related to DI: for example, conveying non-verbal signals to pupils [64], reflecting on teaching performance [6], or monitoring learning progress [8]. These prior design cases imply that distributed peripheral interactive systems are a promising form of technologies to facilitate teachers' DI.

However, the empirical findings of above cases have rarely been connected to the theories of DI [54, 59, 61, 23, 56] in education science, in order to inform future designers of the potential benefits and limitations of distributed peripheral interactive systems. To tackle this unaddressed opportunity, we have explicitly connected our design inquiry to the theories of DI. The aim is to contribute knowledge that contextualizes the DI theories for future design practice, instead of contributing novel design artefacts. Therefore, we use the open-ended platform of FireFlies as a technology probe [33] to contextually explore the potentials and limitations a distributed peripheral interactive system could have in supporting the four elements of DI.

## METHODOLOGY

This study aims to (i) probe the benefits and limitations of distributed peripheral interactive systems in supporting teachers' DI, and (ii) generate design implications and recommendations for future explorations. To this end, we used the FireFlies platform as a probing instrument, to co-design and field-test various interaction solutions with eight secondary school teachers, over the period of six weeks. This section offers an overview of our methodology.

### Participants

We recruited eight teachers from a secondary school in the Netherlands, involving 18 different classes in total. These teachers were intentionally recruited to represent a variety of teaching experiences, from one month to 20 years, because their teaching experience can influence their teaching routines [5]. Furthermore, the participants were teachers of varying disciplines; from native to foreign language and from physical to social science. Details on the participating teachers are elaborated in Table 1.

| #  | Gender | Teaching experience | Age range students | Nr. of lessons |
|----|--------|---------------------|--------------------|----------------|
| P1 | M      | 4 yrs               | 12-19 yrs          | 10             |
| P2 | M      | 20 yrs              | 15-19 yrs          | 6              |
| P3 | W      | 7 yrs               | 16-19 yrs          | 9              |
| P4 | W      | 8 yrs               | 16-19 yrs          | 6              |
| P5 | W      | 20 yrs              | 16-19 yrs          | 9              |
| P6 | W      | 16 yrs              | 12-15 yrs          | 6              |
| P7 | M      | 1 mth               | 13-15 yrs          | 6              |
| P8 | M      | 17 yrs              | 14-17 yrs          | 5              |

**Table 1. Detailed information on the sample of the study.**

### Cyclic process combining co-design and field testing

The six-week study consisted of three cycles of co-design and field test (see Figure 2) in which all teachers participated. Such an repetitive, participatory process was intended to both extract and validate teachers' needs based on rich and in-depth contextual understandings. Moreover, as suggested in [4], what the teachers wanted in the beginning was not always what they really needed. Therefore, engaging teachers in multiple cycles of co-design and field testing allowed us to more accurately sensitize and capture teachers' underlying, and more nuanced needs and experiences. For example, during the first co-design session, the teachers were mostly only envisioning the types of information that students could convey via FireFlies. Yet, having experienced how FireFlies changed the classroom dynamics in the first field trial, they realized that they also needed to ideate proper ways to control when and how the students should interact with FireFlies. We now address the particular methods used in our co-design, field trials, and analysis.

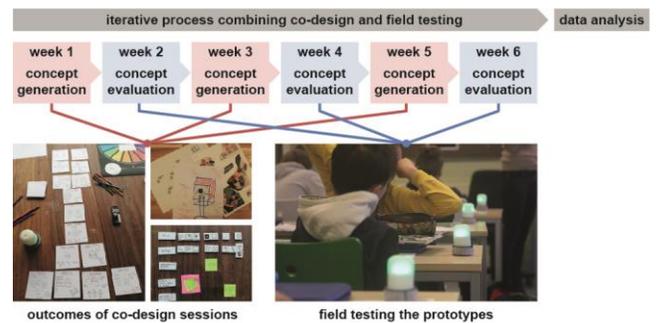

**Figure 2. Overview of the design research methods used**

**Co-design**
Inspired by [32], we utilize co-design as a need-findings approach to broadly explore the opportunities for a peripheral interactive system to support teachers' DI. The co-design approach has been proven effective in developing technologies for teachers, since it ensure that the technologies are aligned with the needs, values, constraints, usefulness and usability in actual classroom contexts [16, 29, 51]. It can be challenging for non-designers (teachers) to meaningfully contribute to the design [39]. Therefore, designing with teachers requires low-threshold generative techniques or strategies [32]. Therefore, in the co-design sessions (Week 1, Week 3, and Week 5), we specifically combined a series of low-threshold generative tools in our toolkit for teachers to express their ideas, including generative card sorting [19], directed storytelling [24], deferred contextual interviews [39] and collage making [5] (see Figure 2). Namely, we ask teachers to freely use these tools to imagine and express their ideas about how they want to teach in the future [49]. These toolkits consisted of visual and written components which were open for interpretation that the teachers could select from to create artefacts that express their thoughts, ideas and dreams which would be difficult to express in words alone [50].

**Field tests of prototypes**
To receive meaningful feedback from the teachers on their concepts, they needed to be aided in understanding the consequences of particular design choices [32]. Prototypes can support this process, as they can enable people to experience a situation that did not exist before [55]. Therefore, during the field test phases (Week 2, Week 4, and Week 6), we deployed prototypes in the teachers' classrooms which embodied the concepts that were generated with the teachers via field observations [27], behavioural mapping [62] and member checking [14].

These prototypes were realized by using the research probe called FireFlies (see Figure 1), introduced by Verweij, Bakker and Eggen [64]. The probe contains a set of wireless interactive ambient lamps: FireFlies. The lamps are controlled by a wireless hub. Each FireFly has a top and a bottom colour. As an open-ended system, the FireFlies lamps can be pre-programmed for different applications. For example, each lamp could be tangibly controlled by a student to change its top colours through simply rotating. A programmable hub could also wirelessly change the colour and animation of the top and bottom of multiple or individual FireFlies.

**Data analysis**
All the discussions during the generative and evaluative phases were audio-recorded and transcribed verbatim and all the observations were described. These qualitative results were analysed through thematic analysis [18] to find (i) empirical knowledge about different areas in which a peripheral interactive system can support teachers in practising DI and (ii) interaction design recommendations to inform HCI design for classroom technology.

**EXAMPLES OF DESIGN OUTCOMES**
Throughout this study, the teachers have generated 20 different concepts. Not all of these concepts were embodied in a prototype and evaluated during the field studies. These concepts were either rejected by the teachers, too difficult to realise or were generated for a certain context which would not occur in the course of the study. Nevertheless, we report these concepts since the conversations about them revealed certain needs of the teachers. In this section we will give some example of use cases that have been tested in the field and concepts that have not been deployed but have led to interesting insights nonetheless.

**Examples of co-design use cases tested in the field**
*Continuous Feedback* allows students to change the colour of their FireFlies on their own initiative and throughout the whole lesson and provide teachers feedback about their levels of understanding or progress.

*Request Feedback* is used when the teacher wants to receive feedback from the students about his/her instruction. Once the teacher turns on all the FireFlies, the students can change their colours to give feedback in a time window. If a student has not changed his/her FireFly to give feedback within this time period, the FireFly will become white.

*Voting system* is used when the teacher want to do a voting with their students about whole-class or individual activities. The interaction in this use case is similar to the interaction in the use case of *Request Feedback*: i.e., students give votes in a time window when the teacher initiates a poll.

*Answering teachers' questions* is used when the teacher wants to see which students feel confident about knowing the answer to a question. The interaction of this use case is similar to that of *Request Feedback*: the teacher initiates a time window and the students indicate their confidence.

*Allow Feedback* is used during the teachers' instruction. When a student has a question about the instruction they can turn their FireFly to a red colour. This change of colour is only visible to the student themselves since all FireFlies are set on a low brightness. When more than five students have a question, the teacher receives a vibrating signal on their phone. In their own time, the teacher can increase the brightness of the FireFlies and reveal which students have a question in order to discuss their questions.

*Request Help* is used when students are working on exercises. When a students has a question, they can turn their FireFly to a red colour. When the student is first in line, their FireFly will have the breathing dimming effect. Once the student no longer needs help, they can turn their FireFly back to green and the effect will stop.

**Examples of co-design concepts that were not deployed in the classroom**
*Show sub-tasks* is used when students are working on a larger assignment that can be divided in sub-tasks. The teachers has colour coded each task in advance and the students will show which task they are working on through their FireFlies.

*Show activity* is used when students can choose among a range of learning activities. By changing the bottom of the FireFlies, the teacher can give advice on what activity the individual students should be working. Likewise, the students can show what activity they are working on with the top of the FireFlies.

*Divide groups* is used when a teacher wants to divide the students in groups based on the level of support that the teachers expect them to need. Through colours, the teacher categorizes the students (e.g. green for no support, yellow for support of their peers and red for extra support from the teacher) and creates homogenic or heterogenic groups accordingly.

*Teachers' expectations* is used for showing the students what level of exercise the teachers think are suitable for them (e.g. green for easy questions, yellow for moderate questions and blue for difficult questions).

**FINDINGS: HOW THE TEACHERS WERE SUPPORTED IN PRACTISING DIFFERENTIATED INSTRUCTION**

The aim of our study is to explore how a peripheral interactive system such as FireFlies can support teachers in DI. The recurrent co-design sessions and field trials over the six weeks have revealed numerous examples of how FireFlies would be, or have been used by the eight participating teachers. Our study thereby has yielded rich and contextualized findings on which aspects of teachers' DI can (cannot) be meaningfully supported with a peripheral interactive system. In this section, we report our findings clustered under the four classroom elements based on the earlier introduced conceptual model of DI: (1) *content*, (2) *process*, (3) *product*, and (4) *learning environment* (see Figure 3 for an overview).

**Differentiating the content**

For the teachers to make informed decisions about when to differentiate the content, the teachers needed feedback on the effectiveness of their instruction. The teachers emphasized the importance of perceiving the readiness during their instruction so they can make on-the-spot alterations [42] to the way they present the information to the students.

*Enhancing the observation of students' body language*

In the classroom, students consciously and unconsciously use smiles, frowns, nodding heads and other nonverbal cues to tell teachers their level of engagement, concentration and understanding [41]. Teachers need to be able to read their students' body language while giving instruction as it gives them feedback on the effectiveness of their instruction. In this study, the teachers expressed that the prototypes helped to increase their awareness of the students' body language. For example, P7 used *continuous feedback* during his instruction and told the students to turn their FireFlies red when they had a question about the content. After his lesson, he said: *"The red lights stand out. When I look in their direction I immediately see when students are no longer listening. In other lessons [without the prototype] I don't notice as quickly that the students have lost their attention."*

Teachers also combined the information from the prototypes with the information from the body language of their students to make sense of their level of engagement. For example, when P1 used the *request help* after his instruction he approached two students who were listening to music and had their heads laid down on their desk. Afterwards, he explained: *"I could see from their bodies that they were not working on their assignment. That would have been fine if they were finished but I could see from their [green] FireFlies that they were not, so I wanted to check what was going on."*

*Receiving up-to-date information about student readiness*

The teachers expressed that they need to receive up-to-date information about their students because the readiness of the student can change multiple times within one lesson. In lessons without the FireFlies, the teachers felt uncertain about their knowledge of their students' readiness. Because of this feeling, the teachers asked their students over and over again whether they understood the instruction. When using *continuous feedback*, this feeling was still present. The teachers explained that they were not sure whether the colours of the FireFlies were up-to-date and that the colours sometimes confused them. P1 described: *"This one time, I thought the students were doing fine because all the FireFlies were green, yet then it turned out that actually many of them had questions but did not think about using their FireFlies. Another time, I saw that a student had turned his FireFly red but when I asked him about it, he said that he did not have a question. The student actually turned his FireFly red a while ago and forgot to change the colour to green after we left that part of the instruction behind us."* By resetting the colours once in a while, as was done with *request feedback*, the teachers were able to receive up-to-date information from their students. Additionally, teachers were less doubtful about the data being up to date, P7: *"Because the colours reset, I know that someone did not accidently show his colour from previous time. The information is up to date for all students."*

*Promoting the students to explicitly think about their level of understanding*

Students are not always aware about their own levels of understanding as it requires mental effort to reflect on it. P2 described: *"I suppose not many of them showed that they had a question, not because they don't dare to or because they are lying, but because it takes a while for them to realise they have a question. Sometimes, students approach me after the lesson or even the day after with a question."* Therefore, teachers need to have their students explicitly thing about their level of understanding during the lesson. With *request feedback* and *answering teachers' questions* the teachers felt that more students actively thought about their levels of understanding. They speculated that the FireFlies helped the students to reflect because of the dedicated moment in which they had to think about any remaining questions.

*Knowing what the students are working on*
Although the teachers did not use the prototypes for this need, the teachers did express their need for an overview of what the students are working on. Especially in classrooms where students were given the freedom to work on an assignment of choice, teachers found it demanding to keep track of each student. P1 envisioned that *show activity* could help him in keeping the overview as it could give him information on which student is working on which assignment in one glance, P1: *"When I want to find out about the different activities that are happening in my classroom, I either ask each individual or I ask students to raise their hand when they are working on assignment A, then again for assignment B, assignment C and so on. It's an inefficient process."* Furthermore, the teachers wanted to know what the students are working on so they can advise them in their learning process. For example, P3 said: *"I like to give students advice on what activity they should do. For instance to let them read the book again if I notice that they still don't understand certain concepts after working on the exercises."*

*Having interactions with a more diverse group of students*
During the idea generation phase in Week 1, the teachers expressed that they were able to form impressions of their students from the interactions that they have with them. The teachers are able to tell who their strongest students are and which students need extra support to reach a learning goal. Yet throughout the study, we observed that in many cases only a small and fixed group of students had interactions with their teacher and that the teacher often only remembered the learning needs of a couple of students.

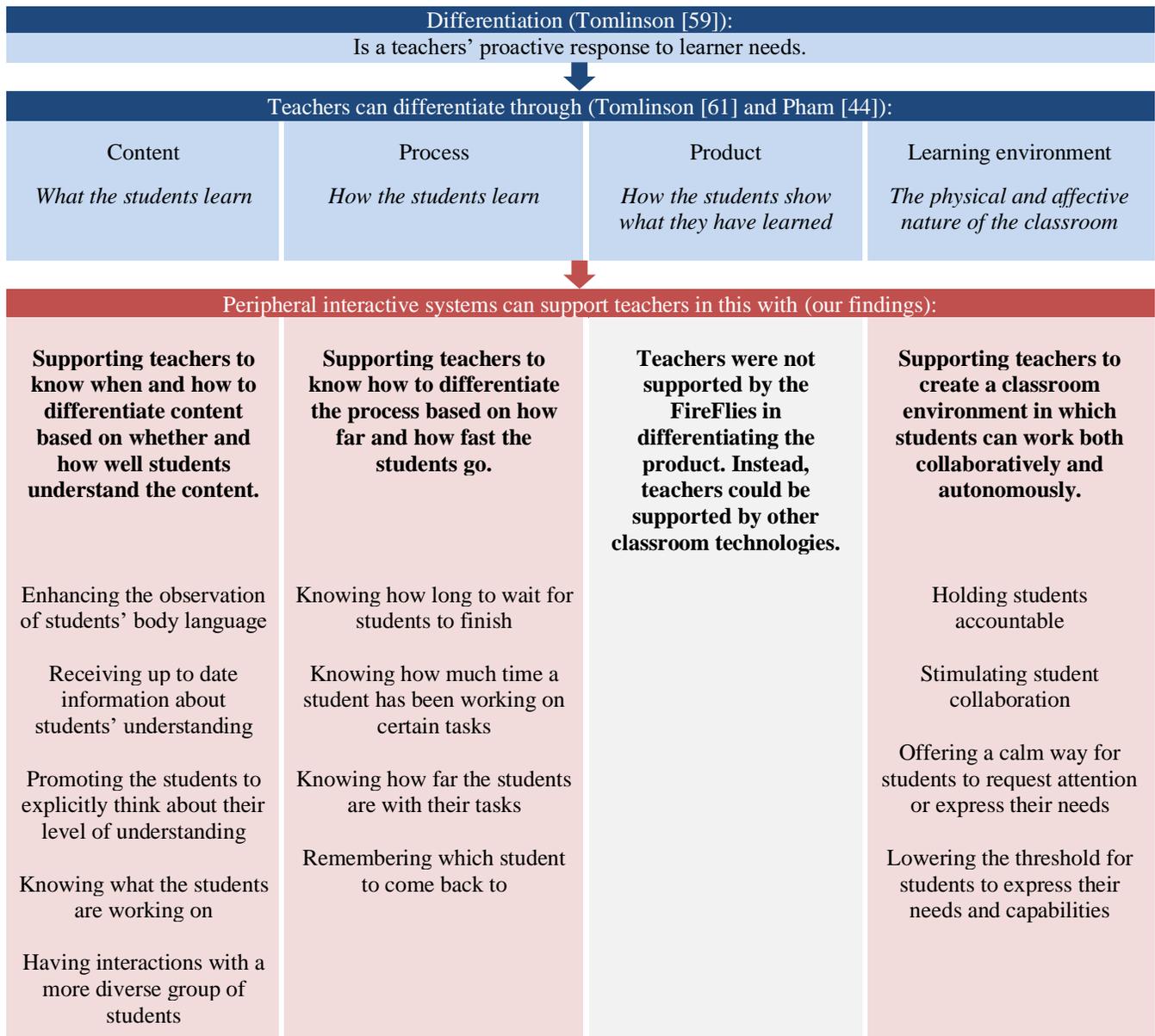

**Figure 3. Conceptual model of Differentiation of Instruction complemented with the ways in which a peripheral interactive system can support teachers in differentiating content, process, product and learning environment.**

The FireFlies supported teachers in having interactions with a more diverse group of students which in turn enabled them to recognize the needs of these students. For example, when P5 used *answering teachers' questions* during a quiz, she told the students to show with their FireFlies if they were able to answer the given question or if they were not. After the quiz, P5 told the observer *"I am surprised about the amount students that didn't know the answer. There were many!"* She explained that in lessons without the FireFlies, there would always be the same group of students who would give answers to the questions and that this might have given her a distorted picture of the readiness of the other students.

**Differentiating the process**
To differentiate the process, the teachers needed an overview of the progress of their students. The FireFlies enhanced this overview and facilitated teachers in offering multiple assignments to the students, giving the students well-timed and personalized support and allowing students to work on their own pace.

*Knowing how long to wait for the students to finish*
When the students were working on short exercises, the teachers needed to know how long they had to wait for the students to finish. The teachers wanted to enable the students to work on their own pace on an individual task before continuing with whole-class instruction. For example, when P5 was using *answering teachers' questions*, she waited until all students had chosen either a red or green colour before appointing a student to answer. After her lesson, she said: *"I liked that I was able to wait until everyone was able to think about the question."* Additionally, the teachers expressed that using the FireFlies to see which students were done with a task reduced their workload, P3: *"Without [the FireFlies] I would also ask the students whether they are done with their task. I like it better to see it in a glance, I no longer have to continuously ask individuals if they are done."*

*Knowing how much time a student has been working on certain tasks*
Although the teachers wanted students to be able to work on their own tempo, they still wanted to be able to stimulate them in their work pace. For this, they needed to know how long it takes for their students to perform certain tasks. For example, P3 wanted to use the FireFlies during a practice exam to coach her students in time management. P3 elaborated: *"[The FireFlies] give me an overview. Now I know this student is working on a certain assignment for a while and I am able to encourage him speed up a little or skip to the next assignment."*

*Knowing how far the students are with their tasks*
When the students were working on longer and more open assignments, the teachers needed to know how far the students were with their tasks. The teachers wanted to offer different type of support throughout different phases of a task. To time this, they need to know in which phase the students are. P1 envisioned that he would like to use s*how sub-tasks* to see when he needs to do check-ups on the quality of the work.

*Remembering which student to come back to*
When the students were working on their exercises, the teachers were walking around the classroom to provide support to the students. We observed that during these moments, multiple students needed the support of their teacher at the same time. Sometimes, the students would have to wait for more than ten minutes before they were able to receive help. Often, students lowered their hand after a while and started to chat with their fellow students or worked on a different assignment. Because of this, the teachers forgot about these students and did not come back to them once they had the time for it. The FireFlies enabled teachers to see which students had a question after they were busy for a while. Because of this, teachers were able to approach students who would otherwise have lost their patience, P4: *"I can see more [when using the FireFlies]. Normally, when the students put their hand up they will get tired after a while and lower it again."* Additionally, *request feedback* supported the teachers in prioritizing which student to help first. P1: *"When I see a blinking FireFly on the other side of the classroom and there is a student near me who also has a question, I ask this student is it is a big or a small question and then I decide who to help first."*

**Differentiating the learning environment**
To differentiate the learning environment, the teachers needed to be able to stimulate collaborative and autonomous learning. Additionally, the teachers needed to create a calm atmosphere in the classroom as they did not feel comfortable to differentiate in a disordered or chaotic classroom.

*Holding students accountable*
The want to teach their students to become more responsible and autonomous. For example, they want their students to make their own decisions about how serious they take their homework and on whether they join classroom instruction or individually practice the exercises. However, as the students are still learning to take these responsibilities, the teachers need to be able to hold their students accountable for their behaviour. When using *voting system,* the FireFlies supported the teachers in this process by informing them about the decisions that each students made. Because of this overview, the teachers could more easily see which students were not sticking to the agreement so they could talk with the students about their responsibilities, P7: *"When a student shows that he want to work autonomously his exercises but I can see that he is talking with other students, I can approach this student and address his behaviour."*

*Stimulating student collaboration*
Some of the teachers expressed that they had difficulties in letting students work in small groups. They explained that the students lack certain collaborative skills because they do not practice collaborative work often enough. Having the students work in small groups therefore required a lot of energy from these teachers. During their lessons, the teachers wanted to let their students practice their collaborative skills through small interactions instead. We observed that the FireFlies played a role in these interactions. For example, P3 was doing a quiz in which the students had to give an answer with groups of four, she used *answering teachers' questions*. The groups were able to higher or lower their stakes in the quiz by turning the FireFlies on or off. During this activity, students used the FireFlies to indicate when they disagreed with their teammates. When student A was writing down the answer without discussing with his teammates, his teammate student B turned off both FireFlies. Only after student A explained his answer to the other two students and these students agreed with the answer, student B turned the FireFlies on again. For another example, when P1 was using *request help*, two students who were seated next to each other showed with their FireFly that they had a question. When P1 approached these students, he asked them if they had a question about the same exercise. When the students turned out to have a question about the same exercise, P1 helped them at the same time and later told them to work on the exercise together.

*Offering a calm way for students to request attention or express their needs*

The teachers, especially of students between 12 and 15 years old, described that the process of students requesting attention can become noisy and disordered. During the observations, we saw that teachers were often walking from one demanding student to the other. The teachers were helping the most notable students first, who were often either loud or close to the location of the teacher. Without the FireFlies, the students tried to catch the attention of their teacher by raising their hand, walking through the classroom, jumping up from their seat or shouting the teachers' name. *Request help* supported the teachers in offering a more calm way for the students to request attention, P1: *"The classroom is calmer [with the FireFlies] because when the students try to get my attention they would [normally] put their hands up and start shouting through the classroom."* Additionally, the students were less distracted while waiting for their teachers' attention, P6: *"When they are holding up their hand, they look around them and start to talk with others and then they are not able to work."*

Furthermore, the teachers expressed that they want to be in control of when to interrupt their instruction, P8: *"When I am mid-explanation and I need to finish my story, I would prefer to wait with the students' questions until I am done."* yet described that in lessons without FireFlies, their students often interrupt them when they have a question. In these lessons, the students' ways of expressing their needs is disturbing the story of the teacher and sometimes students expect the teacher to break of their instruction right when they have a question. P8 explained: *"When I don't respond immediately, which I often can't because I'm still finishing a sentence, they become impatient with me."* *Allow feedback* supported teachers in being in control of when to interrupt their instruction as the FireFlies were set on a low brightness and the students were able to indicate that they had a question without disturbing the flow of the story. Consequently, the teachers were able to decide when they wanted to discuss the questions.

*Lowering the threshold for students to express their needs and capabilities*

The teachers expressed that they highly value their students' feeling of safety. They described that they want their students to be able to be themselves and not be afraid to show their needs or capabilities. The FireFlies helped teachers to lower the threshold for students to express their needs and capabilities. During the field tests, students and teachers experienced that otherwise silent students were now given the turn to ask a question or give an answer to question.

For example, after students of P5 used *answering teachers' questions*, one student told the researcher that he was surprised that one of his fellow students showed his capabilities. He said: *"That student is really smart. But he never raises his hand in class. Yet now he did show that he knew the answer and got a turn to answer a question."* Equivalently, the teachers commented that when they used *request feedback*, they received more information about the needs of students who would not ask questions in lessons without the FireFlies. P2: *"That girl [...] is one of my weakest students. She never asks a question on her own initiative, yet today she did turn her Firefly red."*

**IMPLICATIONS AND RECOMMENDATIONS FOR DESIGN**

Above presented findings reveal both the benefits and the limitations of distributed peripheral interactive systems in supporting different elements of teachers' DI. In this section, we further discuss the findings and formulate design implications and recommendations for future explorations.

Overall, our study suggests that distributed peripheral interactive systems have rich design opportunities to support teachers' DI through three classroom elements: *content*, *process*, and *learning environment*. We have summarized a list of concrete design opportunities for each of the three elements, and some of these insights extend, or confirm the findings of prior explorations. For example, prior work [3, 11, 12, 64] has shown that teachers can seamlessly perceive the information from distributed peripheral system whilst still being able to observe their students. Our results go beyond this by showing that peripheral interactive systems can enhance or optimize teachers' observation of students' body language, by helping them to prioritize which students to look at. In the category of *process*, our teachers felt that the system made their process of delivering help upon students' requests more efficient, which is in accordance with Lantern study [2]. Aligned with [30, 6], we also observed that timely feedback on teaching performance can help teachers reflectively modify their actions on-the-spot.

On the other hand, our study reveals the limitations of distributed peripheral interactive systems. Namely, we found that the classroom element of *product* was not likely to be supported by a system like FireFlies. Instead, this element could be supported by organizational innovations such as customized curriculum and assessment [36], or other forms of technologies such as learning analytics systems [40, 63], intelligent tutoring systems [31, 47, 48] and adaptive assessment systems [37]. Moreover, as implied by this study as well as prior cases, peripheral interactive systems like FireFlies *augment* teachers' DI but do not necessarily *ensure* it. Instead, these systems make DI easier for teachers by lowering the threshold of interpersonal interaction and increasing the accessibility of individual information.

The second aim of our study is to formulate design-oriented insights for future HCI practice aimed for supporting teachers' DI through distributed peripheral interactive systems. We now address these design implications and recommendations, which could be integrated in and further examined by future explorations.

**Complement distributed peripheral information with aggregated display or multimodal notification**
While distributed peripheral information can seamlessly inform teachers in an effortless, unobtrusive manner, in certain moments of time, teachers may have needs for accessing more aggregated, accurate information, or receiving information at a higher notification level. Some examples from our study suggest that it can be meaningful to complement distributed peripheral information with aggregated display, or multimodal notifications. For example, some teachers expressed that they wished to know the amount of students that turned their FireFly to each of the available colours during their instruction. However, when there were more than two colours to choose from, it became challenge for teachers to quickly count the students for each colour. These results suggest that at certain moments, teachers need to access aggregated information in addition to the distributed display.

Higher level notifications through additional or alternative modalities can be sometimes needed to direct teachers' attention to time-relevant signals. We observed that when teachers relied on the FireFlies during their instruction to observe when too many students could not keep up, they often did not notice the amount of students until they finished their instruction. However, our teachers expressed that when too many students can no longer keep up with their instruction, they would like to immediately slow down the pace, give an extra example or change their strategy for the whole classroom. Being able to notice this is important for teachers to decide when to differentiate (e.g. by giving extended instruction to a small group of students after the whole-class instruction) or when to adapt the instruction for all students. In this case, the peripheral display alone is not enough for teachers to shift their focus from their story to the students. Using an additional cue, i.e., vibration from a smart phone, we helped the teachers to perceive the situation while they were focussing on their instruction.

**Show students' self-estimation instead of teachers' assessment to avoid stigmatization in shared display**
Our teachers were very careful to avoid any stigmatisations of their students and therefore rejected ideas of showing their assessment of students on the lamps to group students with same levels. They expected that showing their expectations of the students in the colours of the FireFlies would be confronting for the students: e.g. students who are in the 'weak-group' can feel bad about themselves. On the contrary, concepts that required the students to make estimations about themselves (e.g. level of understanding or progress) were favoured by the teachers. This is because the teachers expected that showing students self-estimation would be less confronting for the students. In addition to that, the teachers preferred to use the FireFlies to request students' self-estimation on their performance in small tasks in a lesson (e.g., "*per questions*"), rather than their overall performance in the whole lesson.

**Regulate the accountability of the teacher to create learner-centred dynamics without pressuring teachers**
While the distributed peripheral information on the lamps could extend teachers' ability of responding to individuals, careful design is needed to avoid increased accountability imposing too much pressure on teachers. One example was observed when P8 was using the earlier version of *allow feedback*. The system revealed that more than five students could no longer keep up with the instruction by increasing the brightness of the FireFlies. When the students did not immediately receive a reaction from the P8 after this revelation, we observed frustrated and angry reactions. Students sighed, or leaned backwards (*"never mind, he won't listen anyway…"*). By contrast, using the later version of *allow feedback*, the teachers were able to regulate when the students can see the colour of the FireFlies of their fellow students. This helped the teachers to avoid their students from feeling being ignored in moments in which the teachers decide not to immediately respond to students' feedback/request. In the deployment of the later version of *allow feedback*, little to none angry or frustrated reactions from the students were observed. These results imply the importance of balancing teachers' enhanced ability with their increased responsibility. In our case, this was done by designing a way to regulate the visibility of student needs and hence to regulate the accountability for teachers to respond.

**Leverage constrains to stimulate interaction or avoid distraction**
Our results suggest that designing constrains (e.g. enabling certain interactions only in specific moments) could stimulate students' interactions with the lamps when their feedback is encouraged, and avoid their distraction when their concentration is needed. For example, when teachers were using *continuous feedback*, we observed that some students were fiddling with their FireFlies. The teachers reported that this could be distracting when they were giving instruction. Furthermore, when teachers wanted the students to change the colour of their FireFly, not all students did so.

This resulted in miscommunication about the colour and sometimes the teachers had to ask individual students if they had turned their FireFly. This problem was resolved by the deployment of *request feedback*, which only enabled students' control of the colour in a specific time window. As observed, when the lamps had the dimming breathing effect to encourage students' feedback, most students immediately responded by turning their FireFlies. Because of the limited time in which the students were able to change their colour, they reacted quickly. Meanwhile, the teachers were less confused about who did and did not turn their FireFly as the FireFlies that were not turned became white. Also, when the FireFlies were disabled, it was no longer a distraction to teachers' instruction, and less students fiddled with it, since it was less entertaining.

**Adopt an ecological lens - consider spatiality of the classroom and enrich existing multimodal interactions**

During this study we noticed that the position of the students and teachers in the classroom influenced the behaviour of the students and teachers.

Firstly, we saw that the teachers often helped students who were seated in close proximity to them earlier than students who were seated further away. When asked about it, they said they were not aware of this pattern, P6: "*I was not aware that I help students who are seated close to me first, it must have been some sort of automated behaviour. I don't really have a strategy for the sequence in which I help students. I think that when someone has a question close to me I help them first because it's efficient but when I know someone else has had a question for a long time I will help them of course.*"

Secondly, as the teachers were walking through the classroom, they were not always able to see all of the FireFlies. Especially when teachers were teaching students aged 16 years and older, they expressed that the bodies of the students were blocking their sight of the FireFlies. In some cases the students responded by repositioning their FireFly until it came in view of their teacher. Based on the teachers' location, the students positioned their FireFly on the right or left side of their desk or positioned it on their neighbours' desk. This behaviour was only seen from students who wanted to catch the attention of their teacher.

Lastly, we saw that the students disregarded the FireFlies when their teacher was already communicating with them, standing next to them or when there were fewer than ten students in the classroom. In these situations, the students used their natural interaction habits such as making eye contact with the teacher, calling the teacher or using their body language to show they have a question.

The teachers envisioned to have students with questions seated near each other so they could help more students at once, or to have students who understands the content help students who do not. However, because of the design of the classroom, these practices caused disturbance in the classroom and some teachers avoided these practices altogether. We believe that, in order to support teachers in aiding this many students, alterations should be made to the classroom design. The teacher needs to be able to group students together on-the-spot and students need to be able to move around without causing disturbance.

**Enable teachers to switch the modes of a distributed peripheral interactive system on the fly**

In Week 4 and Week 6, the teachers were enabled to switch among multiple interaction modes of the FireFlies according to their current need during a class session. We allowed the teachers to do so by using an indication panel next to the whiteboard. However, the teachers rarely used this indication panel and frequently used verbal cues to inform the researcher instead, since the teachers were constantly walking around and had to return to the indication panel to do so. This suggests that teachers should be enabled to change the settings of a peripheral interactive system on the go, e.g. via a remote or wearable control.

**Limitations of the study**

A possible limitation is that some results might be influenced by the novelty effect: although each teacher used FireFlies for 5-10 lessons in total, not all the interaction solutions were deployed throughout these lessons, give our aim of probing user needs instead of evaluating the probe. Thereby, future work could advance our insights through longer or more comprehensive field implementation of the co-design generated ideas. Also, we note that some of our findings could be related to issues of power and control in the classroom, or the 'do-ability' of classroom management, which deserves deepened reasoning in the future. Another relevant direction for future work is to explore the potentials of peripheral interactive systems in terms of widespread adoption and prospective influences in schools.

**CONCLUSION**

In this study, we followed an cyclic co-design process in which we went through three two-week cycles of concept generation and concept validation with secondary school teachers. Through this approach, we were able to gather rich and contextualised data about the usage and experiences with the FireFlies. In this paper, we reported on how a peripheral interactive system was designed through an cyclic co-design process and implemented in three one-week field studies. We found different areas in which a peripheral interactive system can support teachers in practising DI and clustered these areas according to theory about DI, as well as a set of interaction design recommendations for unobtrusively designing HCI for the classroom context.

**ACKNOWLEDGMENTS**
We thank all the participating teachers and students for participating in this study, Dr. S. Bakker for originating the FireFlies platform and Dr. B. Hengevelt and Dr, Y. Chuang for providing valuable feedback in the study, and all our reviewers for insightful suggestions that improved this paper.